\documentclass[draft]{agujournal2019}
\usepackage{url} %this package should fix any errors with URLs in refs.
\usepackage{lineno}
\usepackage{soul}
\usepackage{amsmath}
\newcommand{\Eqref}[1]{Eq.~\eqref{#1}}

\usepackage[normalem]{ulem} 
\draftfalse

\journalname{Geophysical Research Letters}

\begin{document}

%% ------------------------------------------------------------------------ %%
%  Title
%
% (A title should be specific, informative, and brief. Use
% abbreviations only if they are defined in the abstract. Titles that
% start with general keywords then specific terms are optimized in
% searches)
%
%% ------------------------------------------------------------------------ %%

% Example: \title{This is a test title}

\title{Utilizing long memory and circulation patterns for stochastic forecasts of temperature extremes}

%% ------------------------------------------------------------------------ %%
%
%  AUTHORS AND AFFILIATIONS
%
%% ------------------------------------------------------------------------ %%

% Authors are individuals who have significantly contributed to the
% research and preparation of the article. Group authors are allowed, if
% each author in the group is separately identified in an appendix.)

% List authors by first name or initial followed by last name and
% separated by commas. Use \affil{} to number affiliations, and
% \thanks{} for author notes.
% Additional author notes should be indicated with \thanks{} (for
% example, for current addresses).

% Example: \authors{A. B. Author\affil{1}\thanks{Current address, Antartica}, B. C. Author\affil{2,3}, and D. E.
% Author\affil{3,4}\thanks{Also funded by Monsanto.}}

\authors{Johannes A. Kassel and Holger Kantz}

% \affiliation{1}{First Affiliation}
% \affiliation{2}{Second Affiliation}
% \affiliation{3}{Third Affiliation}
% \affiliation{4}{Fourth Affiliation}

\affiliation{1}{Max Planck Institute for the Physics of Complex Systems, Nöthnitzer Straße 38, 01187 Dresden, Germany}
%(repeat as many times as is necessary)

%% Corresponding Author:
% Corresponding author mailing address and e-mail address:

% (include name and email addresses of the corresponding author.  More
% than one corresponding author is allowed in this LaTeX file and for
% publication; but only one corresponding author is allowed in our
% editorial system.)

% Example: \correspondingauthor{First and Last Name}{email@address.edu}

\correspondingauthor{Johannes A. Kassel}{jkassel@pks.mpg.de}

%% Keypoints, final entry on title page.

%  List up to three key points (at least one is required)
%  Key Points summarize the main points and conclusions of the article
%  Each must be 140 characters or fewer with no special characters or punctuation and must be complete sentences

% Example:
% \begin{keypoints}
% \item	List up to three key points (at least one is required)
% \item	Key Points summarize the main points and conclusions of the article
% \item	Each must be 140 characters or fewer with no special characters or punctuation and must be complete sentences
% \end{keypoints}

\begin{keypoints}
\item We present a data-driven modeling technique for one-dimensional nonlinear stochastic models with long memory and external driving
\item Causal influence of NAO and AO indices on daily minimum and maximum temperature anomalies in Europe is strongest in southern Scandinavia and extends beyond two weeks
\item Including long memory and exogenous driving by the AO index significantly enhances forecast horizon in one-dimensional stochastic forecasts of 2m maximum and minimum temperature anomalies in European winter
\end{keypoints}

%% ------------------------------------------------------------------------ %%
%
%  ABSTRACT and PLAIN LANGUAGE SUMMARY
%
% A good Abstract will begin with a short description of the problem
% being addressed, briefly describe the new data or analyses, then
% briefly states the main conclusion(s) and how they are supported and
% uncertainties.

% The Plain Language Summary should be written for a broad audience,
% including journalists and the science-interested public, that will not have 
% a background in your field.
%
% A Plain Language Summary is required in GRL, JGR: Planets, JGR: Biogeosciences,
% JGR: Oceans, G-Cubed, Reviews of Geophysics, and JAMES.
% see http://sharingscience.agu.org/creating-plain-language-summary/)
%
%% ------------------------------------------------------------------------ %%

%% \begin{abstract} starts the second page

\begin{abstract}
Long memory and circulation patterns are potential sources of subseasonal-to-seasonal predictions. Here, we infer one-dimensional nonlinear stochastic models of daily temperature which capture both long memory and external driving by the Arctic Oscillation (AO) index. To this end, we employ a data-driven method which combines fractional calculus and stochastic difference equations. A causal analysis of AO and North-Atlantic Oscillation indices and European daily extreme temperatures reveals the largest influence of the AO index on winter temperature in southern Scandinavia. Stochastic temperature forecasts for Visby Flygplats, Sweden, show significantly improved performance for long memory models. Binary temperature forecasts show predictive power for up to 20 (11) days lead time for maximum (minimum) daily temperature (66\% CI) while an AR(1) model possesses predictive power for 8 (3) days lead time for daily maximum (minimum) temperature (66\% CI). Our results show the potential of long memory and circulation patterns for extreme temperature forecasts.
\end{abstract}

\section*{Plain Language Summary}
% Topic overview
Extending the forecast horizon of weather models, in particular for extreme event prediction, is crucial for preventing human and economic loss. Potential sources for improving weather forecasts include using slowly decaying correlations of temperature fluctuations, also called long memory, and long-lived atmospheric pressure patterns which characterize macro weather. Additionally, new data-driven modeling techniques may incorporate these sources and complement traditional numerical weather models based on physical equations.
% Paper overview
In this study, we introduce a modeling technique which is able to learn interpretable phenomenological models involving randomness from time series data. It incorporates the long memory of temperature fluctuations as well as external driving by atmospheric pressure patterns. We then use models reconstructed from temperature data recorded at a Swedish weather station to make temperature predictions and assess their performance.
% Paper findings
We find that the dominant atmospheric pressure patterns in the northern hemisphere influence European maximum and minimum temperature fluctuations significantly for more than two weeks. 
Although our learned models involving long memory and atmospheric pressure patterns are drastically simpler than a weather forecast, they are useful for up to 20 days which is not the case for baseline models incorporating randomness with neither long memory nor macro weather information. The additional driving by an atmospheric pressure pattern improves the forecast at small forecast times.
% Key takeaways
Our work shows the impact of including long memory in simple data-driven weather models involving randomness, motivating further research to improve forecast performance.

%In this study, we build a simple model which includes random effects and also has the long memory feature. Additionally, our model incorporates the macro weather situation, which has a longer life span than the local weather. As we show, the macro weather situation influences local extreme winter temperatures in Scandinavia for more than two weeks. We then use our model to predict extreme winter temperatures for one weather station in Sweden.

%% ------------------------------------------------------------------------ %%
%
%  TEXT
%
%% ------------------------------------------------------------------------ %%

%%% Suggested section heads:
\section{Introduction}
Long memory, also called long-range dependence or long-range correlations, is characteristic of many geophysical data sets, e.g. temperature anomalies, river run-off data and precipitation time series \cite{Hurst1951,Koscielny-Bunde1996,Eichner2003,Fraedrich2003,Bartos2005,Kiraly2006, Bartos2006}. These fractal structures in time are present on the daily, monthly, and yearly time scale \cite{Rybski2008,Franzke2020}. While the origins of long-range dependence and the length of the memory remain debated \cite{Mann2011,Maraun2004,Fredriksen2017}, we assume long memory to be a given feature of the data. Among the consequences of long-range dependence are enlarged confidence intervals for temporal averages and trends compared to time series without memory \cite{Rybski2006,Ko2008,Lennartz2009,Lennartz2011,Tamazian2015,Phillips2023}. Here, the inherent persistence of the geophysical phenomena impedes the estimation of quantities.

One conceptual framework for understanding long-range correlations is the climate memory approach \cite{Yuan2013,Yuan2014,Yuan2019,Nian2020}. Via fractional differencing, a long-range correlated geophysical time series may be decomposed into the short-term weather component and the cumulative memory component which represents the climate system \cite{Yuan2013}. Utilizing this decomposition, several authors proposed and quantified the potential of long memory to improve predictions of geophysical phenomena \cite{Yuan2019, Nian2020}.

Another source of predictability for atmospheric dynamics are circulation patterns \cite{White2017, Nian2020, Domeisen2022}. They have significant impact on atmospheric variability across wide geographical areas. Their longevity, providing a link between weather and climate \cite{Barnston1987}, makes their inclusion in weather models a natural candidate to improve forecast performance and push the prediction horizon into the subseasonal-to-seasonal range, i.e. 10 to 30 days \cite{White2017, Nian2020, Domeisen2022}. The North-Atlantic Oscillation (NAO) and the Arctic Oscillation (AO) are the dominant circulation patterns for the European continent. Both AO and NAO are known to significantly influence surface air temperature across the European continent, especially extreme events \cite{Hurrell1996, Thompson2001, Scaife2008, Pozo2001}. A physical explanation lies in AO and NAO variability's correspondence to the location of the winter polar jet stream in the stratosphere of the northern hemisphere \cite{Hurrell2003}. These changes dictate if winds reaching Europe originate from the Arctic or the Atlantic Ocean, hence transporting cold or mild air masses. Both indices are available on the daily scale.

Many recent stochastic modeling techniques used in geophysics either incorporate nonlinearities but lack long memory \cite{Hasselmann1976,Franzke2015a,Palmer2019,Franzke2020} or incorporate long memory but lack nonlinearities \cite{Mandelbrot1968,Mandelbrot1968noah,Hosking1981,Granger1980,Graves2017}. Here, we use an approach by \citeA{Kassel2022} which is apt to infer stochastic models including nonlinearities and long memory. The method uses fractional integration to introduce long memory and thus is compatible with the climate memory approach mentioned above. We extend the approach to include an additional external forcing by a circulation mode index, assuming a unidirectional coupling between circulation mode and surface air temperature.

The remainder of this Letter is structured as follows. At first, we investigate the influence of atmospheric circulation patterns, namely the North-Atlantic Oscillation (NAO) and the Arctic Oscillation (AO) on daily minimum and maximum temperature anomalies in European winter (DJF). Secondly, we reconstruct a nonlinear, long-range correlated stochastic model of daily maximum and minimum temperature anomalies from temperature data recorded at Visby, Flygplats (SE) driven by the AO index. Finally, we employ the reconstructed model to forecast daily maximum and minimum temperature in winter. To assess forecast performance, we compare the model forecast to stochastic model forecasts which possess neither nonlinearities nor long memory.

\section{Circulation Modes and Extreme Temperature}
We use the lagged Pearson correlation coefficient and the lagged mutual information to spatially resolve varying influences of the AO on maximum and minimum daily temperature anomalies in European winter. Introducing a time lag $\tau$ between the two time series, we measure the strength of the correlation as a function of $\tau$ \cite{Wilks2011}:
\begin{align}
r_{X,Y}(\tau) = \frac{\mathrm{Cov}[X,Y](\tau)}{\sigma_X \, \sigma_Y} &= \frac{\left\langle \, (X(t) -\langle X\rangle) (Y(t-\tau) - \langle Y\rangle) \, \right\rangle}{\sqrt{\left\langle X^2 - {\langle X\rangle}^2 \right\rangle \left\langle Y^2 - {\langle Y\rangle}^2\right\rangle}} \,.
\end{align}
For a bivariate random variable $ Z = (X, Y)$, the mutual information $I(X, Y)$ measures how the uncertainty in one random variable is reduced by the knowledge about the other random variable \cite{Gray2011}. In contrast to the (linear) correlation, it is zero if and only if $X$ and $Y$ are independent.

Analogous to the lagged Pearson correlation coefficient, we introduce a time lag $\tau$ between the time series $X$ and $Y$:
\begin{align}
I(X(t), Y(t-\tau)) = H(X_+) + H(Y_-) - H(X_+, Y_-) \,,
\end{align}
in which $H$ denotes the Shannon entropy and the indices $+, -$ indicate the shortening of the time series $X_+ = \{X_i\}_{i\in[\tau+1,N]}$, $Y_- = \{Y_i\}_{i\in[1,N-\tau]}$ with $N$ the length of the respective time series. %The lagged mutual information measures how much uncertainty about $X$ is reduced by $Y$ lagged by $\tau$ time steps. 
The standard estimator for the mutual information $I(X, Y)$ is the $k$-nearest neighbor algorithm introduced by \cite{Kraskov2004}. The free parameter $k$ determines how many nearest neighbors should be included for the estimation. Larger values of $k$ reduce statistical fluctuations but lead to a negative bias of the estimator \cite{Kraskov2004}.

We use ERA5 reanalysis data generated by the European Centre for Medium-Range Weather Forecasts (ECMWF), provided via their Copernicus platform \cite{Thepaut2018}. The original ERA5 data set covers the time window 1979--2022 \cite{Hersbach2020}, with an extension covering 1950--1978 \cite{Bell2021}.
For the AO index, we use daily time series provided by the Climate Prediction Center of the National Oceanic and Atmospheric Administration (NOAA) \cite{CPC2022}.
We conducted the analysis presented here also with weather station data from the ECAD data set \cite{ECAD} and obtained similar results. However, due to the inhomogeneous spatial distribution of weather stations, the figures obtained with the reanalysis data are more demonstrative and easily interpretable.

We extract daily minimum and maximum surface temperatures for every grid point from the hourly reanalysis data, resulting in daily maximum and minimum temperature time series from January 1st, 1950 until March 31st, 2022. Subsequently, we approximate the seasonal cycle at every grid point by a second-order Fourier series and subtract it to obtain the temperature anomalies. For each grid point we compute the lagged mutual information and the lagged Pearson correlation coefficient for the meteorological winter, December until February (DJF). We compute the lagged mutual information using the \emph{pyunicorn} package implementation \cite{Donges2015} of the Kraskov I algorithm and use $k=0.03 \, N$ nearest neighbors in order to minimize estimator bias and statistical fluctuations \cite{Kraskov2004}.
\begin{figure}
    \centering
    \includegraphics[width=1.0\textwidth]{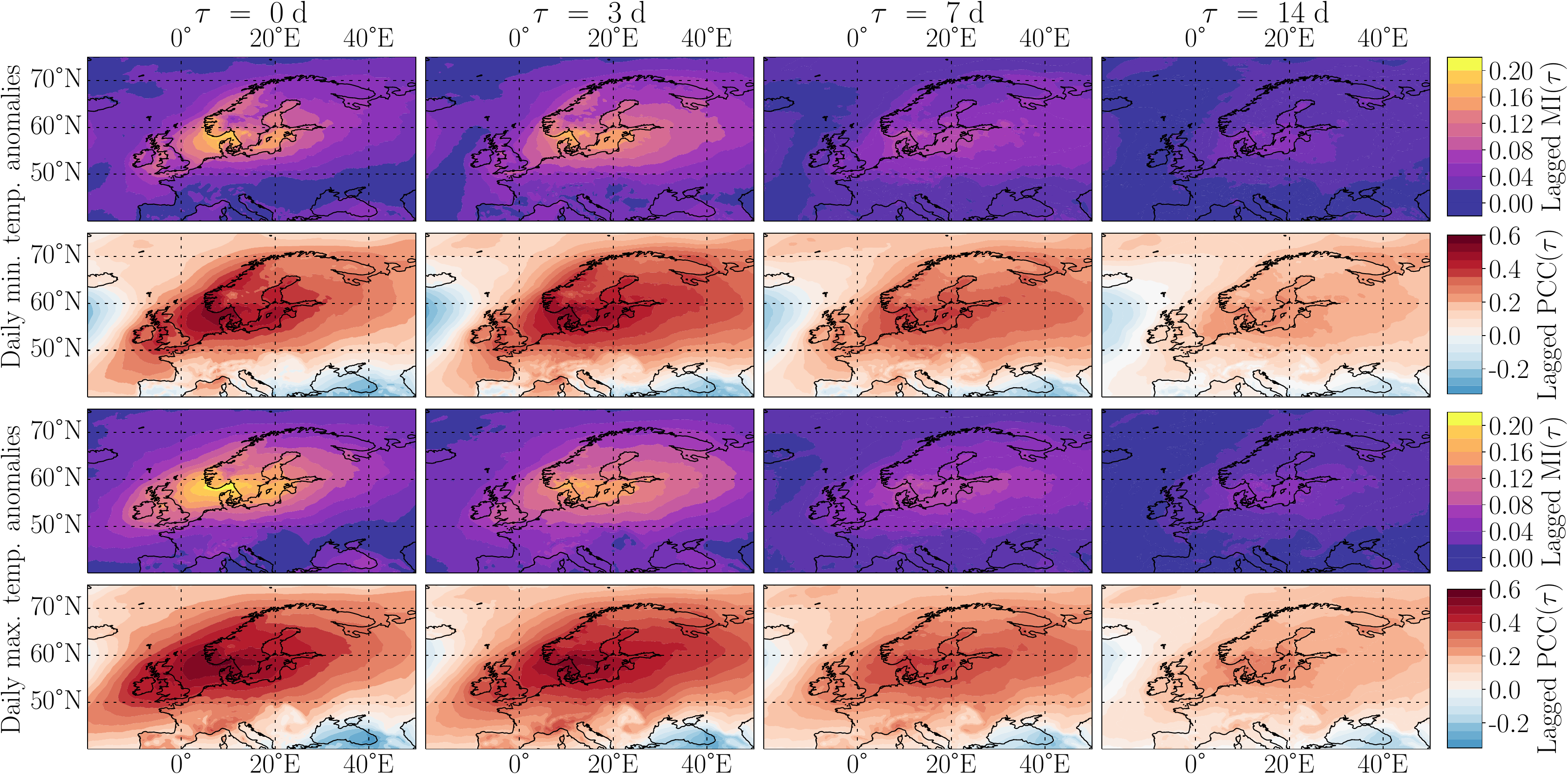}
    \caption{Lagged mutual information (Lagged MI) and lagged Pearson correlation coefficient (Lagged PCC) of the AO index and daily maximum and minimum temperature anomalies in winter (DJF). Contour lines show elliptical shape with the maximum in southern Scandinavia. Lagged PCC remains above $r>0.3$ in southern Scandinavia up to 14 days lag time.}
    \label{fig:causal_analysis_era5}
\end{figure}
As Figure~\ref{fig:causal_analysis_era5} shows, the AO shows strongest influence on surface temperature anomalies in southern Scandinavia, reaching correlation values of $r=0.6$ ($r=0.55$) for daily maximum (minimum) anomalies at two days lag time. %TMIN TMAX
The influence isolines show elliptical shape which have larger meridional extension for the daily maximum temperature case than for the daily minimum anomalies. This is also reflected by the zero correlation contour line which is located further North for the daily minimum anomalies.
The ellipse of strong positive influence slightly shifts eastward as the time lag increases. The correlation values weaken for increasing time lag but remain above $r=0.25$ for both daily maximum and minimum daily temperature anomalies in southern Scandinavia at two weeks lag time. The lagged mutual information figures confirm these findings, suggesting a linear relationship between the AO index and daily temperature anomalies.

\begin{figure}
    \centering
    \includegraphics[width=0.49\textwidth]{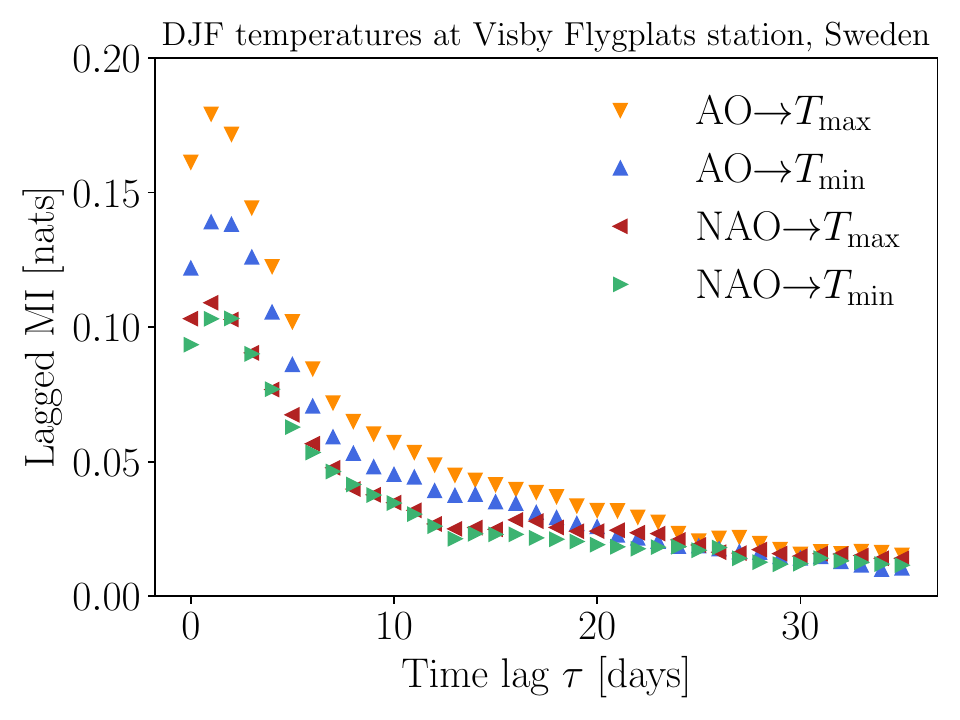}
    \includegraphics[width=0.49\textwidth]{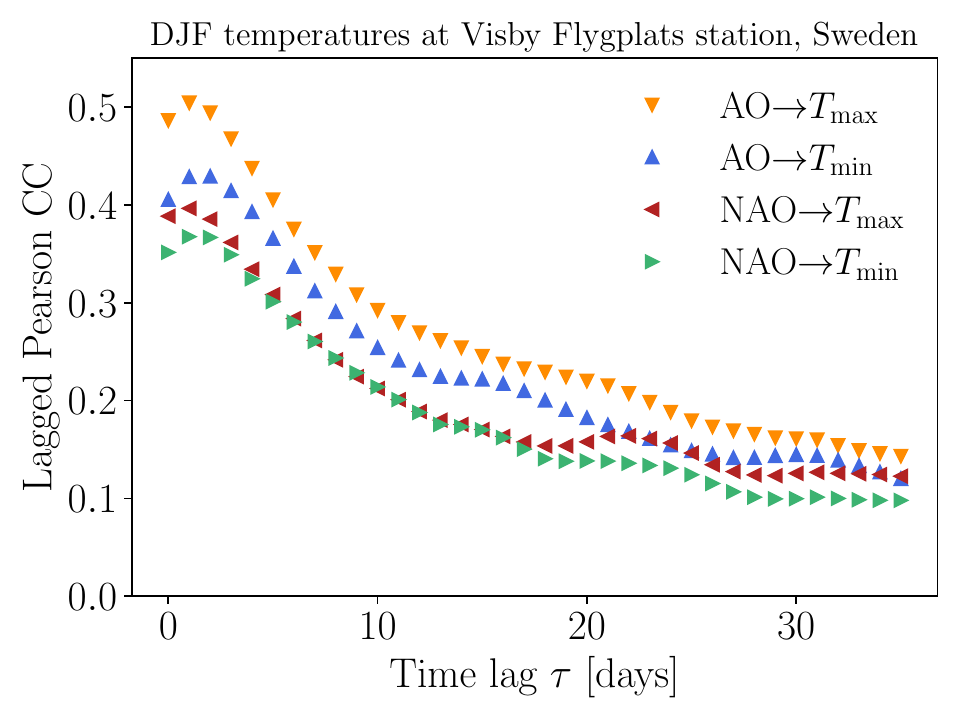}
    \caption{\textbf{Causal Analysis of AO and NAO index and winter daily extreme temperatures recorded at Visby Flygplats, Sweden.} Lagged mutual information (left panel) and lagged Pearson correlation coefficient (right panel). None of the curves shows zero-crossings even after five weeks, suggesting a long-lasting influence of the large-scale pressure patterns associated with AO and NAO indices on daily extreme temperature anomalies. Influence on daily maximum anomalies is larger than on daily minimum anomalies for both indices while AO's influence on temperature anomalies is larger than NAO's. The findings match the causal analysis of reanalysis data (cf. Fig.~\ref{fig:causal_analysis_era5}).}
    \label{fig:causal_visby}
\end{figure}

We search the ECAD data set \cite{ECAD} for a weather station located in southern Scandinavia whose records cover at least the time period from 1950 to 2022 without missing data points because the AO index is available from 1950 onward. This leaves us with the Visby Flygplats weather station located at the coast of the island Gotland, Sweden, in the Baltic Sea. Figure~\ref{fig:causal_visby} shows the repeated causal analysis for the NAO and AO indices and the daily minimum and maximum temperature anomalies of the Visby Flygplats station data.
All curves possess a maximum after one or two days and show a consecutive decay which flattens after approximately one week. The influence of the AO is larger than the NAO's influence at all time lags. The influence of the AO on the daily maximum temperature anomalies is the largest among all investigated index-temperature couplings, surpassing $r=0.5$ at one and two days lag time. These results agree well with the reanalysis data presented above. We therefore proceed to reconstruct a stochastic model of the temperature time series recorded at the Visby Flygplats station and employ it for predictions.

\section{Nonlinear Long-Range Correlated Stochastic Models with External Driving}
The inference method we use here is an extension of the method introduced in \citeA{Kassel2022}. The method is based on the assumption of the separability of the short-range and long-range dynamics implying that the Hurst exponent is accessible independently of the short-range dynamics of the process (cf. \cite{Hosking1984}). In a two-step procedure, we first estimate the Hurst exponent of the temperature anomalies and then apply a fractional filtering to the temperature anomalies to remove the long memory. Afterwards, we employ a maximum-likelihood approach to estimate a Markovian stochastic difference equation of the fractionally differenced temperature anomalies and the AO index time series as an exogenous driver. We estimate the Hurst exponent of the temperature anomalies using  DFA-3, i.e. detrended fluctuation analysis with a third-order polynomial \cite{Peng1994,Hoell2019} and obtain $H = 0.71$ for the Visby daily minimum temperature anomalies and $H = 0.70$ for the Visby daily maximum temperature anomalies. The subsequent fractional filtering is motivated by the connection between the fractional integration in ARFIMA processes \cite{Hosking1981,Granger1980} and the Gr\"unwald-Letnikov fractional integral \cite{Podlubny1998}. This connection proposes the application of the inverse operation, the first-order difference approximation of the Gr\"unwald-Letnikov fractional derivative, to a time series $T'_n$ in order to remove long-range correlations:
\begin{align}
T_n = {}_{n-M}^{\quad \mathrm{GL}}\Delta^{d}_{n} \; T'_n = \sum_{j=0}^M \frac{\Gamma(j-d)}{\Gamma(j+1) \Gamma(-d)} \; T'_{n-j} \,,
\end{align}
in which $M$ is the memory length, $d$ the memory parameter with $d=H-1/2$ and $\Gamma(\cdot)$ the Gamma function. Since the fractional differencing is a filter eliminating $M$ data points, choosing $M$ is a trade-off between memory removal and data sparsity. Here, we set $M=5\;{\mathrm{years}}$ which is also due to a criterion for the memory length $M$ based on the tolerable error of the fractional differencing (see SI).

The resulting time-series $T_n$ is short-range correlated but not necessarily Markovian. However, based on the exponential decay of the autocorrelation function, the Hurst exponent amounting to $H\approx 0.5$ estimated via DFA, a passed Chapman-Kolmogorov test and partial autocorrelation functions vanishing after one time lag we consider the fractionally differenced temperature anomalies Markovian.
Hence, we infer a Markovian stochastic difference equation of the following form:
\begin{align}
T_{n+1} &= f(T_n, y_{n-\tau}) + g(T_n, y_{n-\tau}) \, \xi_{n+1} \,,\label{eq:discrete_langevin_exogenous_driving}
\end{align}
in which the deterministic drift term $f$ as well as the stochastic diffusion term $g$ may depend on both the fractionally differenced temperature anomaly $T_n$ and the AO index $y_{n-\tau}$ lagged by $\tau$ days. Exploiting the information about large-scale weather patterns provided by the AO, we expect better forecast performance for the driven model than for the pure temperature model.
Here, we use $f(T_n, y_{n-\tau};\{\lambda\}) = \lambda_4 T_n^3 + \lambda_3 T_n^2 + \lambda_2 T_n + \lambda_1 y_{n-\tau} + \lambda_0$ and $g^2(T_n;\{\theta\}) = \theta_4 T_n^4 + \theta_3 T_n^3 + \theta_2 T_n^2 + \theta_1 T_n + \theta_0$.
We also infer models with nonlinear dependence of the deterministic force on the AO index as well as a dependence of the diffusion term on the AO index but they do not improve forecast performance. The noise $\xi$ is Gaussian white noise with zero mean and variance one.

For model estimation, we restrict the fractionally differenced temperature anomalies to the winter months (DJF). Thereafter, we divide this ensemble of time series into training set and test set. Starting with 1955, every fourth winter is in the test set, resulting in $N_{\mathrm{test}}=18$ test set members. Three quarters of the fractionally differenced winter temperature anomalies are used for the estimation of Markovian models. We concatenate these time series to obtain the training data set, but exclude data point pairs overlapping two different winters from the estimation procedure. For a detailed description of the maximum likelihood estimation scheme, see SI.

As baseline models, we use the persistence $T^{\prime}_{n+1} = T^{\prime}_n$ and the Markovian AR(1) process $T^{\prime}_{n+1} = \phi \, T^{\prime}_n + \sigma \, \xi_{n+1}$. To assess the impact of nonlinearities and the forcing by the AO index, we also use the ARFIMA(1,d,0) process $T^{\prime}_{n} = \phi \, T^{\prime}_{n-1} + {}_{n-M}^{\quad \mathrm{GL}}\Delta^{d}_{n} \, \xi_n$. Finally, we also infer a bivariate linear model from the fractionally differenced temperature anomalies $T_{n+1} = \phi_1 \, T_n + \phi_2 \, y_{n-\tau} + \sigma \, \xi_{n+1}$.

\section{Stochastic Forecast}
Having obtained the optimal parameters $\{\hat{\lambda}\}$ and $\{\hat{\theta}\}$, we can generate trajectories employing the following stochastic difference equation with the acquired parameters:
\begin{align}
T_{n+1} = f\left(T_n, y_{n-\tau};\{\hat{\lambda}\}\right) + \sqrt{g^2\left(T_n; \{\hat{\theta}\}\right)} \; \xi_t \,,\label{eq:discrete_langevin_model}
\end{align}
where $\xi_t$ is Gaussian white noise with zero mean and unit variance.
By construction, time series generated using Eq.~\eqref{eq:discrete_langevin_model} are Markovian and have similar statistical properties as the fractionally differenced time series. Then, we employ a fractional integration exactly as in ARFIMA processes, which is the inverse operation of the fractional differencing applied beforehand,
\begin{align}
    T^{\prime}_{n} = {}_{n-M}^{\quad\mathrm{GL}}\Delta^{-d}_{n} \; T_n = \sum_{j=0}^M \frac{\Gamma(j+d)}{\Gamma(j+1) \Gamma(d)} \; T_{n-j} \,,
\end{align}
in which once again the memory parameter $d$ is determined by the estimated Hurst parameter of the temperature anomalies and $M$ has the same value as before. This step introduces long-range correlations and thus models interactions with the climate system \cite{Yuan2013}.

For every winter in the test set, we perform forecasts starting at dates from December 1st to February 27th. For each start date, we generate an ensemble of $10^4$ trajectories whose initial condition is the fractionally differenced temperature anomaly on the start date $t_0$ of the forecast. Each of these trajectories has a length of at most $35$ days but does not exceed February 28th. We transform the Markovian trajectories into temperature predictions by fractionally integrating the generated trajectory concatenated with the fractionally differenced temperature anomalies ($M$ and $d$ as before) and addition of the seasonal cycle. Finally, we calculate the trajectory ensemble mean for each day of the forecast but also save the histogram of the trajectory ensemble on each forecast day.

The models including atmospheric forcing incorporate the AO index time series lagged by $\tau$ days. For a causal prediction, $\tau$ cannot remain constant throughout the iteration as this would require AO index realizations after the prediction start date $t_0$. We therefore estimate a family of models with $\tau \in \{\tau_{\mathrm{min}}, \tau_{\mathrm{min}} + 1, \ldots, \tau_{\mathrm{max}} -1, \tau_{\mathrm{max}}\}$. For the daily maximum temperatures, the minimum lag is $\tau_{\mathrm{min}}=1\;\mathrm{day}$ and for the daily minimum temperatures $\tau_{\mathrm{min}}=2\;\mathrm{days}$, since the AO influence is maximal for these respective lag times (cf. Fig.~\ref{fig:causal_visby}). During the forecast, after $\tau_{\mathrm{min}}$ time steps, in each forecast step $i\in\{1,\ldots,\tau_{\mathrm{max}}-\tau_{\mathrm{min}}\}$ we employ the next model estimated for $\tau_{\mathrm{min}} + i$ days lag time, i.e. a lag time increased by one day compared to the previous forecast step.

\section{Forecast Results}
\begin{figure}
\centering
\includegraphics[width=0.49\textwidth]{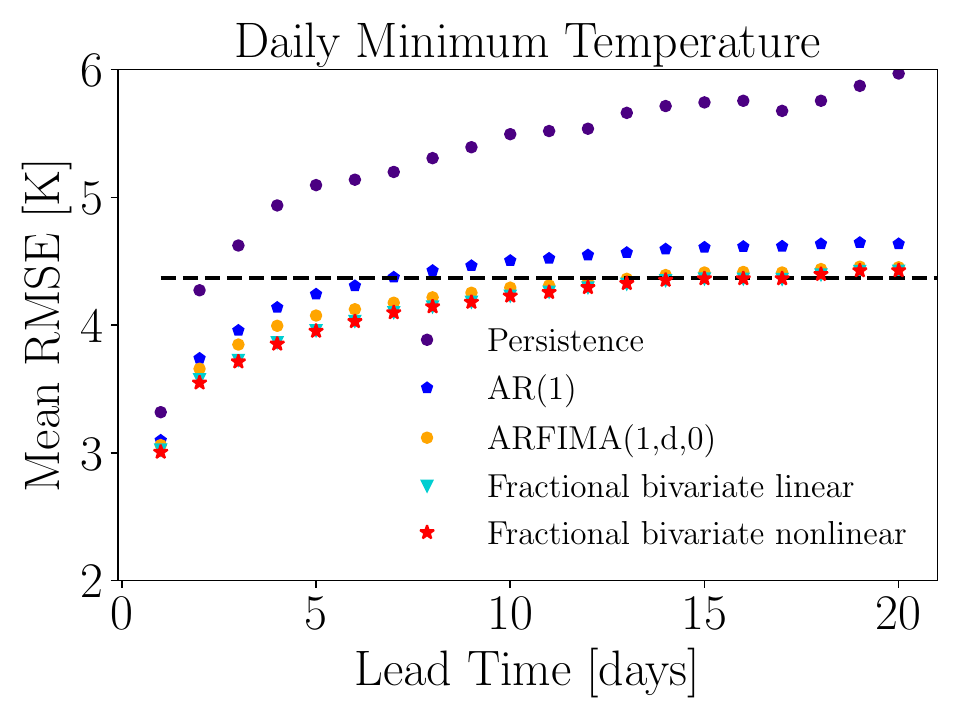}
\includegraphics[width=0.49\textwidth]{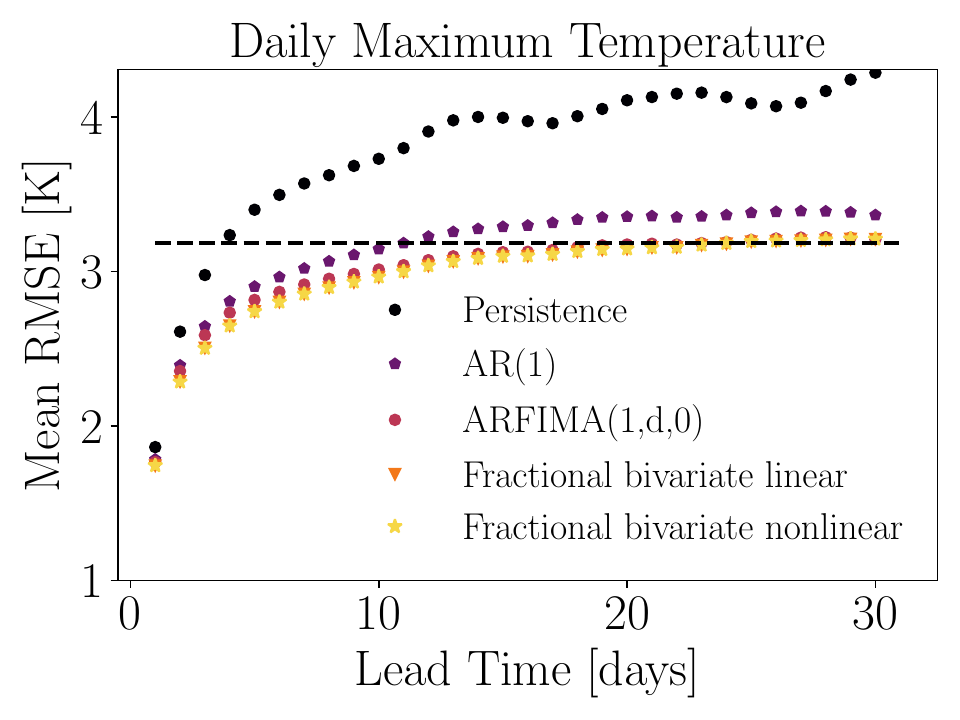}
\caption{\textbf{Mean RMSE of winter daily temperature forecasts for Visby Flygplats, Sweden.} Fractional models significantly outperform Markovian models. \textbf{Left panel: Daily minimum temperature forecasts.} %2,6,13,17,17
The AR(1) model has predictive power for up to six days and the persistence for two days lead time, after which the standard deviation of the observed temperatures (black dashed line) is crossed. The ARFIMA(1,d,0) has predictive power for up to $12$ days while both models incorporating the AO index have the same RMSE and cross the standard deviation of observed temperatures after $14$ days lead time. \textbf{Right panel: Daily maximum temperature forecasts.} The AR(1) has predictive power for $11$ days lead time while the persistence has predictive power for three days. The ARFIMA(1,d,0) RMSE is larger than the RMSE of the AO index models at smaller lead times but crosses the standard deviation of observed temperatures as well after $23$ days.} %3,11,24,24,24
\label{fig:rmse_markov_fractional}
\end{figure}

We use the root-mean-square error (RMSE) and the Brier skill score (BSS) to assess the performance of the forecast \cite{Wilks2011}. Averaging over the test set and all forecast start dates $\{\mathrm{SD}\}$, the RMSE reads
\begin{align}
\mathrm{RMSE}(t) = \frac{1}{|\{\mathrm{SD}\}|} \sum_{t_0\in\{\mathrm{SD}\}}\left[\frac{1}{N_{\mathrm{test}}} \, \sum_{i=1}^{N_{\mathrm{test}}} \left(X_i(t_0+t) - O_i(t_0+t)\right)^2\right]^{1/2} \,.
\end{align}
Here, $X_i(t_0 + t)$ is the predicted temperature for the test set member $i$ with forecast start date $s$ after $t$ prediction steps. $O_i(s+t)$ then is the observed minimum or maximum temperature on that day. 
We compare the RMSE of the model forecasts to the standard deviation of the winter minimum and maximum temperatures in the test set, shown by Figure~\ref{fig:rmse_markov_fractional}.
In general, the RMSE of the fractional models' forecasts crosses the dotted black line at later lead times than the RMSE of the Markovian models. Thus, including LRC in stochastic temperature models significantly increases forecast performance. Furthermore, we observe that incorporating the external forcing by the AO index decreases forecast RMSE, especially at smaller lead times. This forecast improvement is more pronounced for daily minimum temperature anomalies than for daily maximum temperature anomalies. Here, the nonlinear model's RMSE does not differ from the RMSE of the bivariate linear model. The predictive power of the daily minimum temperature forecast increases from six days for the AR(1) forecast to $14$ days for the fractional models driven by the AO index. The predictive power of the daily maximum temperature forecast increases from $11$ days for the AR(1) forecast to $23$ days for the fractional models driven by the AO index.

\begin{figure}
\centering
\includegraphics[width=0.495\textwidth]{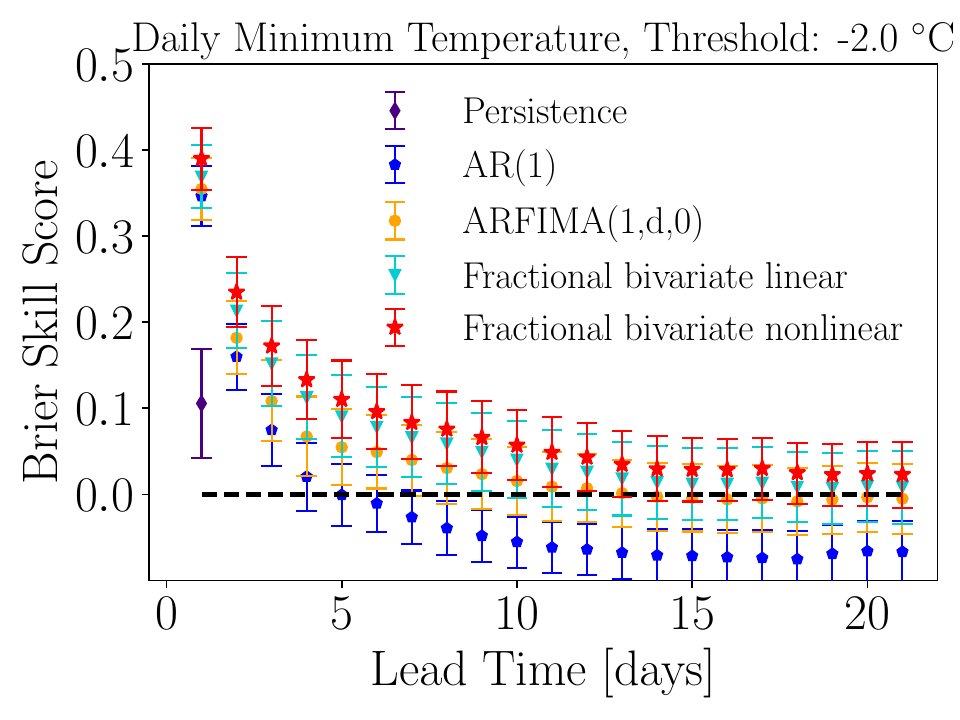}
\includegraphics[width=0.495\textwidth]{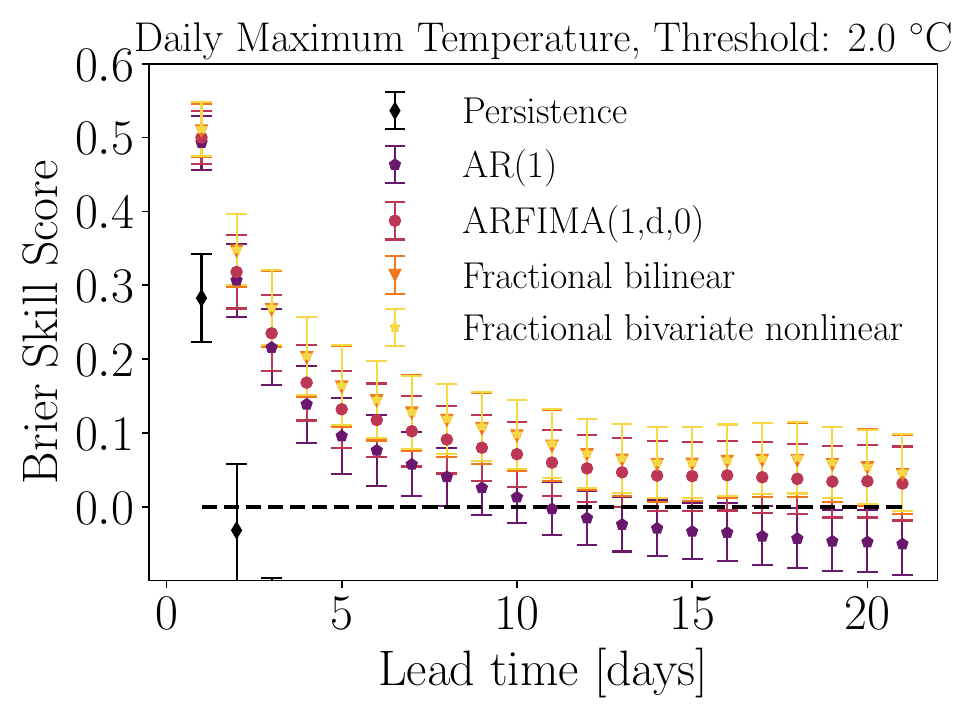}
\caption{\textbf{Brier skill scores of binary threshold crossings for winter daily minimum temperature forecast at Visby Flygplats, Sweden. Left panel:}
The AR(1) model has predictive power of three days lead time and the persistence for one day lead time. Models driven by the AO index perform better than the AR(1) and ARFIMA(1,d,0) models.  \textbf{Right panel:} Forecasts with models driven by the AO index show higher BSS than the AR(1) and ARFIMA(1,d,0) models. Both models incorporating the AO index perform identically. The AR(1) model has predictive power for up to eight days lead time and the persistence for one day. Bootstrapped error bars amount to $66\%$ confidence intervals \cite{Wilks2011}.}
\label{fig:bss}
\end{figure}

Secondly, we assess the binary forecast skill, i.e. the ability of the model to predict threshold crossing, which measures the ability of the forecast to predict extremal temperatures. The BSS is a popular measure to assess the performance of binary forecasts relative to a reference forecast using the Brier score (BS) \cite{Brier1950,Wilks2011}. It is defined as \cite{Wilks2011}
\begin{align}
\overline{\mathrm{BSS}}(t) &= 1 - \frac{1}{|\{\mathrm{SD}\}|} \sum_{t_0\in \{\mathrm{SD}\}} \frac{\mathrm{BS}(t, t_0)}{\mathrm{BS}_{\mathrm{ref}}(t,t_0)} \,, \label{eq:marginalized_brier_skill_score}\\
\mathrm{BS}(t, t_0) &= \frac{1}{N_{\mathrm{test}}} \sum_{i=1}^{N_{\mathrm{test}}} \left(p_i(t, t_0) - O^{\mathrm{thresh}}_i(t_0+t)\right)^2 \label{eq:brier_score} \,.
\end{align}
Here, \Eqref{eq:marginalized_brier_skill_score} defines the Brier skill score marginalized over all start dates of the forecast. \Eqref{eq:brier_score} defines the Brier score in which $p_i(t, t_0)$ denotes the predicted threshold crossing probability of the test set member $i$ at lead time $t>t_0$ and start date $t_0$ which lies in the set of start dates $\{\mathrm{SD}\}$ with cardinality $|\{\mathrm{SD}\}|=89$. The symbol $O^{\mathrm{thresh}}_i(t_0+t)$ represents the observed threshold crossing, thus $O^{\mathrm{thresh}}_i(t_0+t)\in\{0, 1\}$.
We use the climatological frequency of threshold crossing in the test set as the reference forecast, as common in the literature \cite{Wilks2011}. We obtain the predicted threshold crossing probability from the histogram of the forecast trajectories ensemble. %A Brier skill score above zero indicates that the prediction has better skill than the climatological frequency \cite{Wilks2011}.
Figure~\ref{fig:bss} shows the Brier skill scores for daily maximum and minimum temperature forecasts and one threshold each with $66\%$ confidence intervals.
Fractional models perform significantly better than Markovian models, showing predictive skill for up to $20$ ($12$) days lead time for daily maximum (minimum) temperatures. In contrast, the AR(1) forecast has predictive power of eight (three) days lead time for daily maximum (minimum) temperatures.

Remarkably, the nonlinear model of daily minimum temperatures performs better than the bivariate linear one, which differs from the forecast RMSE and the results for daily maximum temperatures. While the forecast RMSE only takes the mean of the forecast trajectory ensemble into account, the threshold crossing probability depends on the histogram of the ensemble. The histogram of daily maximum temperature anomalies is well approximated by a Gaussian, meaning linear models are sufficient. This coincides with the equivalent performance of the bivariate linear model with the nonlinear model.
Daily minimum temperature anomalies, however, strongly deviate from Gaussianity (see SI), rendering linear models inappropriate. Nonlinear models perform better at reproducing the non-Gaussian tails which are relevant for the threshold crossing probabilities.

% results)
%
\section{Summary and Conclusion}
We have presented a data-driven method to infer one-dimensional nonlinear stochastic models with long memory and external driving. It possesses a time-scale separation which allows for a physical interpretation. The short-time Markovian dynamics models weather fluctuations while the the long-time dynamics models couplings with the climate system. The latter is introduced via a fractional integration also used in ARFIMA models. This is in line with the climate memory approach \cite{Yuan2013, Yuan2014, Yuan2019}.

We have investigated the influence of the Arctic Oscillation on winter daily minimum and maximum temperature anomalies as a function of the lag time and found a maximum area of influence in southern Scandinavia, where the Pearson correlation coefficient surpasses $r=0.5$ for both maximum and minimum temperature anomalies at two days lag time and remains above $r=0.25$ even after two weeks lag time.

We have inferred stochastic models with long memory for daily minimum and maximum temperature anomalies recorded at Visby Flygplats weather station in Sweden for winter (DJF). Both forecast RMSE and Brier skill score of binary forecasts show significantly improved forecast performance for models with long memory, bearing predictive power of up to $20$ days for daily maximum temperatures. Forcing by the AO index improves forecast skill especially at small lead times.

\section{Open Research}
The daily maximum and minimum temperature reanalysis data used for the causal analysis are available via the Copernicus platform of ECMWF at \url{https://cds.climate.copernicus.eu/datasets/} \cite{Hersbach2020,Bell2021}.
The daily Arctic Oscillation index \url{https://ftp.cpc.ncep.noaa.gov/cwlinks/norm.daily.ao.cdas.z1000.19500101_current.csv} and North-Atlantic Oscillation index time series \url{https://ftp.cpc.ncep.noaa.gov/cwlinks/norm.daily.nao.cdas.z500.19500101_current.csv} are available at the Climate Prediction Center webpage of the National Oceanic and Atmospheric Administration \cite{CPC2022}.
The Visby weather station data used for the causal analysis as well as the temperature forecast are available at European Climate Assessment \& Dataset via \url{https://www.ecad.eu/} \cite{ECAD, Besselaar2015}.
Version 0.6.1 of the pyunicorn package used for computing the mutual information is available via \url{https://www.pik-potsdam.de/~donges/pyunicorn/index.html} and developed openly at \url{https://github.com/pik-copan/pyunicorn} \cite{Donges2015}.
The routines for computing the causal analysis as well as the model inference, temperature forecasts and forecast analysis as well as plotting routines are available at \url{https://zenodo.org/uploads/14597489} \cite{zenodo}.

%%%%%%%%%%%%%%%%%%%%%%%%%%%%%%%%%%%%%%%%%%%%%%%

%% ------------------------------------------------------------------------ %%
%% References and Citations

%%%%%%%%%%%%%%%%%%%%%%%%%%%%%%%%%%%%%%%%%%%%%%%
%
\bibliography{winter_extreme_temp}
%
% don't specify bibliographystyle

% In the References section, cite the data/software described in the Availability Statement (this includes primary and processed data used for your research). For details on data/software citation as well as examples, see the Data & Software Citation section of the Data & Software for Authors guidance
% https://www.agu.org/Publish-with-AGU/Publish/Author-Resources/Data-and-Software-for-Authors#citation

%%%%%%%%%%%%%%%%%%%%%%%%%%%%%%%%%%%%%%%%%%%%%%%

%\bibliography{enter your bibtex bibliography filename here}

%Reference citation instructions and examples:
%
% Please use ONLY \cite and \citeA for reference citations.
% \cite for parenthetical references
% ...as shown in recent studies (Simpson et al., 2019)
% \citeA for in-text citations
% ...Simpson et al. (2019) have shown...
%
%
%...as shown by \citeA{jskilby}.
%...as shown by \citeA{lewin76}, \citeA{carson86}, \citeA{bartoldy02}, and \citeA{rinaldi03}.
%...has been shown \cite{jskilbye}.
%...has been shown \cite{lewin76,carson86,bartoldy02,rinaldi03}.
%... \cite <i.e.>[]{lewin76,carson86,bartoldy02,rinaldi03}.
%...has been shown by \cite <e.g.,>[and others]{lewin76}.
%
% apacite uses < > for prenotes and [ ] for postnotes
% DO NOT use other cite commands (e.g., \citet, \citep, \citeyear, \citealp, etc.).
% \nocite is okay to use to add references from your Supporting Information
%

\end{document}

\maketitle
	
	%\begin{abstract}
	%	This file contains supporting information for the letter ...
	%	
	%	\noindent\textbf{Keywords:} article, template, simple
	%\end{abstract}

	\tableofcontents

\section{Influence of North-Atlantic Oscillation on extreme winter temperature in Europe}
\begin{figure}
    \centering
    \includegraphics[width=1.0\linewidth]{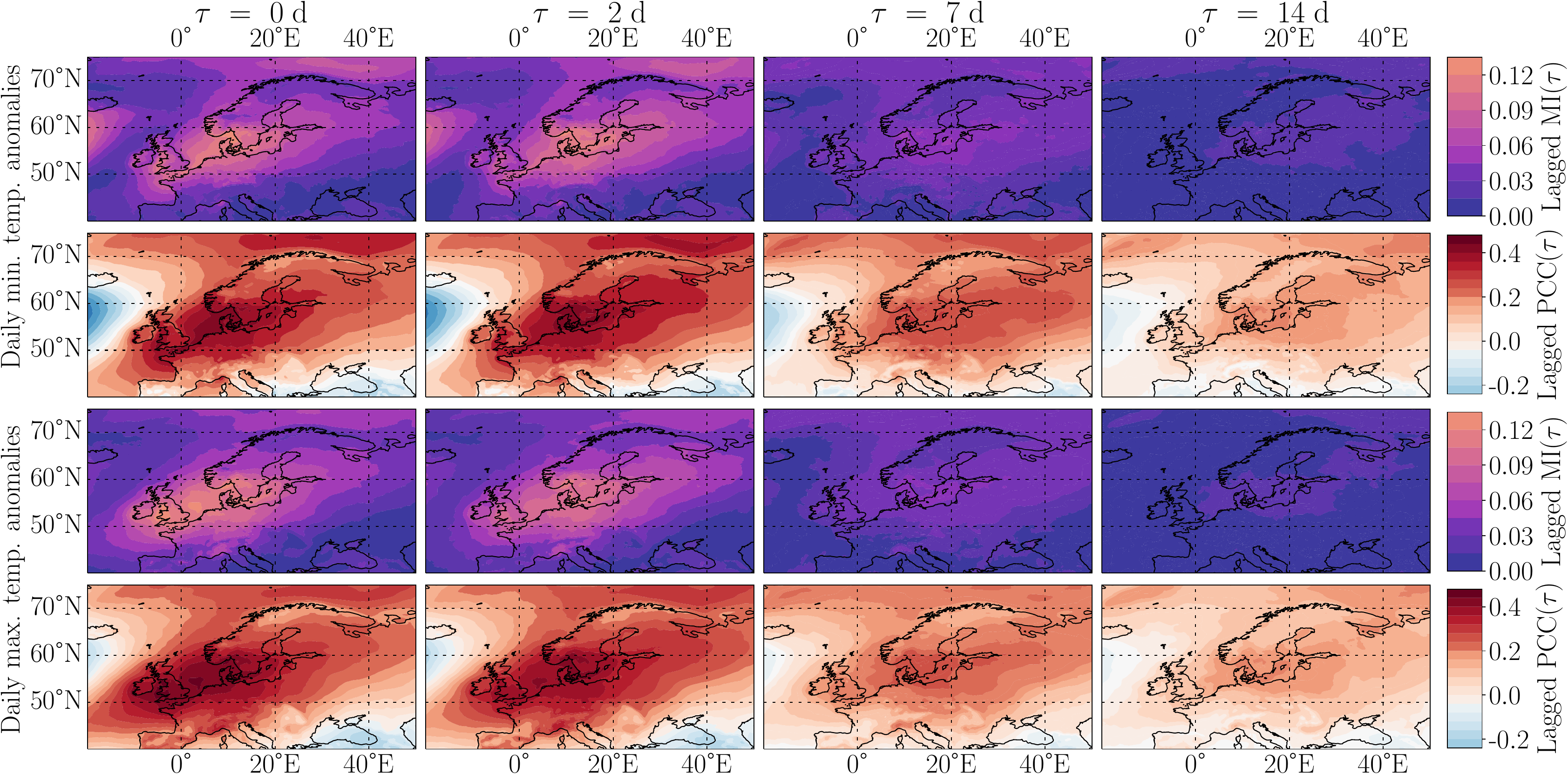}
    \caption{Lagged mutual information (Lagged MI) and lagged Pearson correlation coefficient (Lagged PCC) of the NAO index and daily maximum and minimum temperature anomalies. Contour lines show elliptical shape with the maximum in southern Scandinavia. Lagged PCC remains above $r>0.2$ in southern Scandinavia up to 14 days lag time.}
    \label{fig:causal_NAO}
\end{figure}

Figure~\ref{fig:causal_NAO} shows the influence of the North-Atlantic Oscillation (NAO) index on daily maximum and minimum temperature anomalies in European winter (DJF). It is obtained in the same manner as the figure showing the AO influence on temperature anomalies presented in the main text.
In comparison to the influence patterns of the Arctic Oscillation index (see main text), the isolines show a larger westward extension and include the British Isles. They shift eastwards for increasing lag time and remain significant even after two weeks lag time as for the Arctic Oscillation index. The magnitude of the lagged Pearson correlation coefficient does not exceed $r=0.45$ and hence is lower than the influence of the Arctic Oscillation index.

\section{Estimation of the Seasonal Cycle}
At first, we remove February 29th from each leap year of the temperature time series to ensure that all years of the time series have the same length. This does not impede further analysis because it amounts to less than $1$\textperthousand \, data points of the time series. Subsequently, we estimate the seasonal cycle $S(t_i)$, also referred to as climatology. To this end, we employ a second-order Fourier series with a basic angular frequency of $\omega=\frac{2\pi}{365 \; \mathrm{d}}$, as defined by the following formula:
\begin{align}
S(t_i) = c + a_1 \cos\left(\frac{2 \pi}{365 \; \mathrm{d}} \, t_i\right) + b_1 \sin\left(\frac{2 \pi}{365 \; \mathrm{d}} \, t_i\right) + a_2 \cos\left(\frac{4 \pi}{365 \; \mathrm{d}} \, t_i\right) + b_2 \sin\left(\frac{4 \pi}{365 \; \mathrm{d}} \, t_i\right) \,. \label{eq:deseasonalisation}
\end{align}
The coefficients are determined with a linear regression. Upon removal of the seasonal cycle from the daily maximum (minimum) temperatures, we obtain the daily maximum (minimum) temperature anomalies $\{T_{\mathrm{max}}\}$ ($\{T_{\mathrm{min}}\}$. These are the deviations from the long-term mean modeled by the seasonal cycle. By definition, the daily maximum and minimum temperature anomalies are an approximately stationary time series with mean zero.

We ignore slow-mode variations as well as the global warming trend in the time series. Estimating the local warming trend based on one time series is challenging and also complicated by the long-range correlations of the anomalies \cite{Rybski2006, Ko2008, Lennartz2009, Lennartz2011, Tamazian2015}. Subtracting an assumed trend would introduce new artifacts to the time series while the validity of the local warming trend's elimination remains unclear. Taking the global warming trend as a proxy, it amounts to approximately $1\;\mathrm{K}/100 \; y$ whereas the standard deviation of the daily mean temperature anomalies amounts to $\sigma = 3.9\;\mathrm{K}$. Hence, the global warming trend only marginally violates the stationarity of the daily time series. 

\section{Memory Selection for Fractional Differencing}
In the fractional calculus literature, the error resulting in computing a fractional derivative with truncated memory is given by the `short-memory' principle \cite{Podlubny1998}. For a deterministic bounded function $f(t) \leq C$, the error $e(t)$ stemming from approximating a fractional derivative by a shorter memory length is \cite{Podlubny1998}
\begin{align}
e(t) = |_a^{\mathrm{GL}}D_t^d f(t) - {_{t-M}^{\mathrm{GL}}D_t^d} f(t)| \leq \frac{C \,  M^{-d}}{|\Gamma(1-d)|} \,,
\end{align}
with $a+M\leq t$. Here, $_a^{\mathrm{GL}}D_t^d$ is the Gr\"unwald-Letnikov fractional derivative with memory parameter $d$ and terminals $a$ and $t$. For larger memory length $M$, the upper bound of the error decreases according to a power law determined by $d$. The short-memory principle also allows to determine the memory length necessary to achieve a given accuracy \cite{Podlubny1998}.

\begin{figure}
\centering
\includegraphics[width=0.495\textwidth]{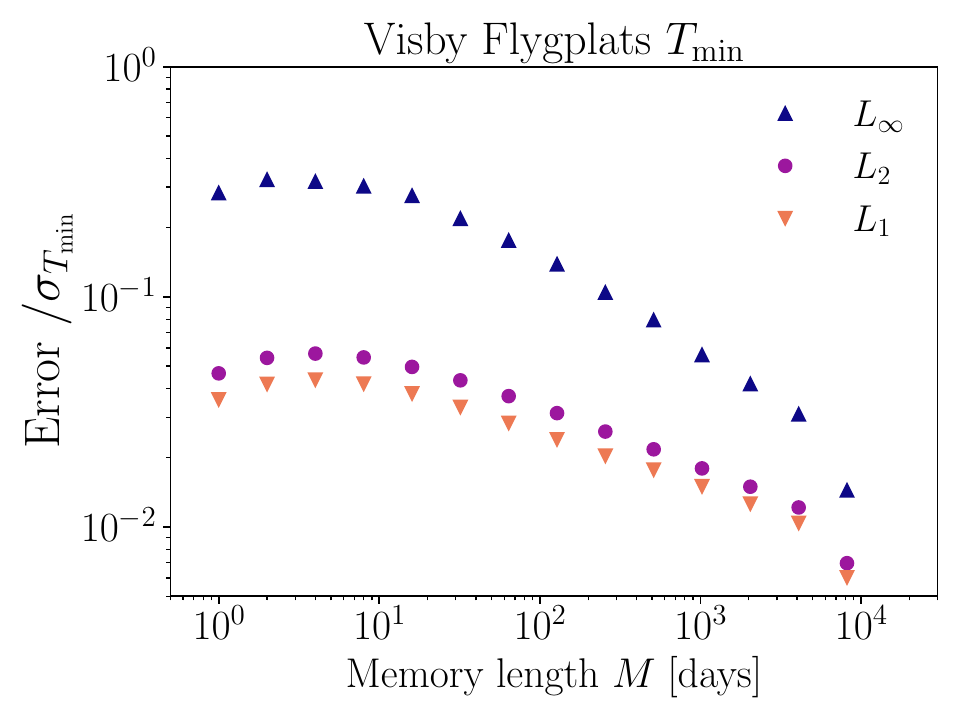}
\includegraphics[width=0.495\textwidth]{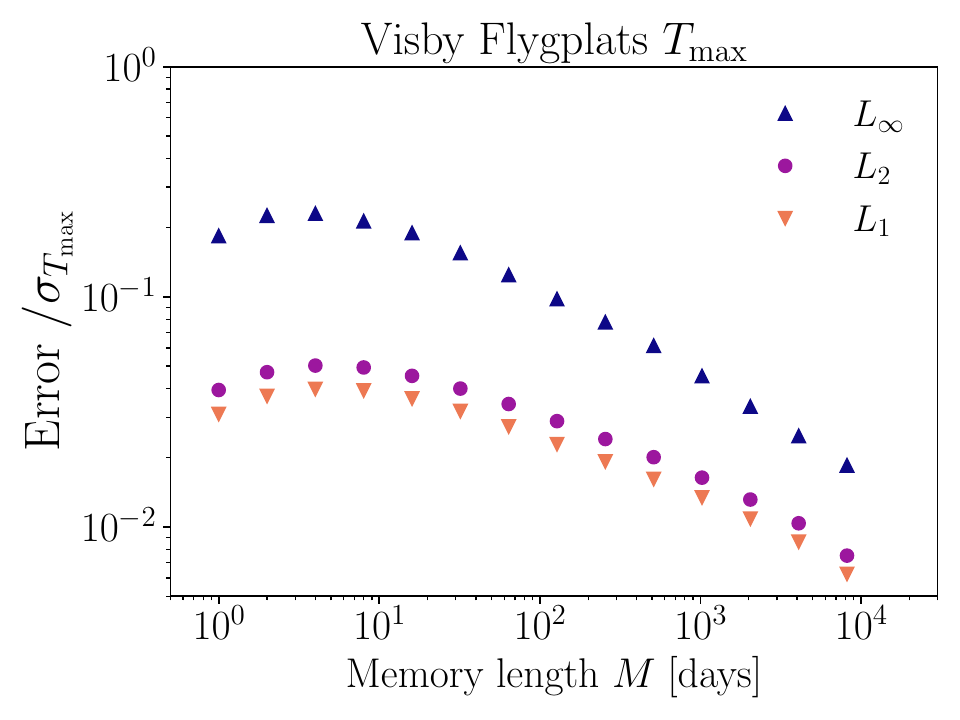}
\caption{\textbf{Fractional differencing errors of daily extreme temperature anomalies
recorded at Visby Flygplats, Sweden. Left panel:} Normalized errors of memory selection criterion for daily minimum temperature anomalies. The $L_{\infty}$ error falls below $0.1 \; \sigma_{T_{\mathrm{min}}}$ at $M \geq 512 \; \mathrm{days}$ while the $L_2$ error falls below $0.01\;\sigma_{T_{\mathrm{min}}}$ at $M = 8096 \; \mathrm{days}$. \textbf{Right panel:} Normalized errors of memory selection criterion for daily maximum temperature anomalies. The $L_{\infty}$ error falls below $0.1\;\sigma_{T_{\mathrm{max}}}$ already at $M = 128 \; \mathrm{days}$ and the $L_2$ error falls below $0.01 \; \sigma_{T_{\mathrm{max}}}$ at $M = 8096 \; \mathrm{days}$.}
\label{fig:visby_memory_selection}
\end{figure}

Since we do not know the `true' values of the fractional derivative of real-world time series, we propose the following memory selection criterion. For various memory lengths $M$, we first apply a fractional derivative with $d=\hat{H} - 0.5$ and a subsequent fractional integration of the fractionally differenced time series with the same $M$ and $d$. Since the two operations are inverse, the resulting time series should coincide with the original time series, of course omitting the first $2M$ data points lost due to the filtering. We then calculate the p-norm of the error for the remaining $N-2M$ data points and divide it by the standard deviation of the original time series:
\begin{align}
\hat{e}(M) = \frac{||_{t-M}\Delta_t^{-d} \{_{t-M}\Delta_t^d \{y_t\}\}  - y_t||_p}{\sigma_y} \,,\label{eq:memory_selection_criterion}
\end{align}
in which the fractional differencing is the first-order difference approximation of the Gr\"unwald-Letnikov fractional derivative, defined as follows \cite{Podlubny1998}:
\begin{align}
_{t-M}\Delta_t^{d} y_t = \tau^{-d} \sum_{j=0}^M \frac{\Gamma(j-d)}{\Gamma(j+1)\Gamma(-d)} \, y_{t-j} ~\overset{\tau = 1.0}{=}~ \sum_{j=0}^M \frac{\Gamma(j-d)}{\Gamma(j+1)\Gamma(-d)} \, y_{t-j} \,.
\end{align}
Here, we set the sampling interval, i.e. the finite time difference, to zero.
Dividing by the standard deviation renders the error dimensionless and makes it comparable.
This scaled error may serve as an criterion for selecting the appropriate memory length. It also allows to obtain a quantitative estimate of the errors for a chosen memory length $M$. The selection of an error tolerance threshold $e$ then determines the required memory length $M$. Figure~\ref{fig:visby_memory_selection} shows the evaluation of the error criterion for daily minimum and maximum temperature anomalies recorded at Visby Flugplats, Sweden. For model estimation and forecast we choose $M = 5 \; \mathrm{years}$.

\section{Histogram of Visby Flygplats Temperature Anomalies}
\begin{figure}
\centering
\includegraphics[width=0.49\textwidth]{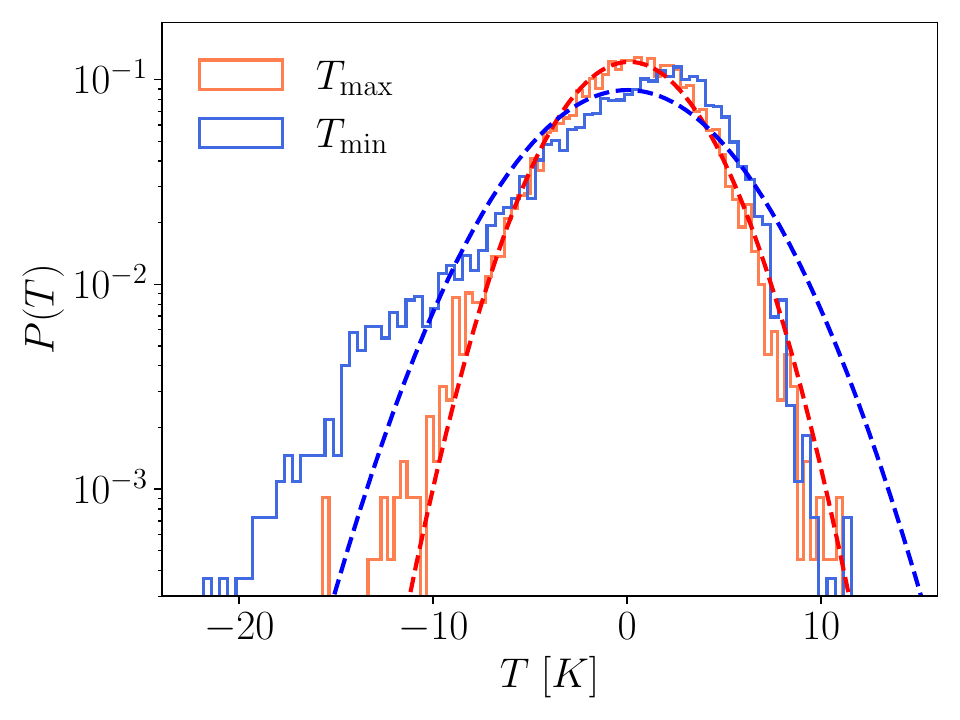}
\caption{\textbf{Histograms and Gaussian fits of daily maximum and minimum winter (DJF) temperature anomalies recorded at Visby Flygplats, Sweden}. The daily maximum temperature anomalies are approximately Gaussian, while the daily minimum temperature histogram is strongly skewed, deviating from Gaussianity with an approximately exponential tail for negative anomalies. Hence, large negative daily minimum temperature anomalies are more likely than positive daily minimum temperature anomalies of the same magnitude.}
\label{fig:winter_temp_anomalies_visby_hist}
\end{figure}

Figure~\ref{fig:winter_temp_anomalies_visby_hist} shows the histograms of daily minimum and maximum temperature anomalies in winter (DJF) recorded at Visby Flugplats, Sweden together with Gaussian fits. The winter daily maximum anomalies are well approximated by the normal distribution while the winter daily minimum anomalies are strongly skewed and thus strongly deviate from the normal distribution.

\section{Estimation of the Stochastic Difference Equation}
We employ a stochastic difference equation to model the fractionally differenced temperature anomalies which arises naturally from the Euler-Maruyama scheme:
\begin{align}
T_{n+1} &= T_n + \Delta t \tilde{f}(T_n, y_{n-\tau}) + \sqrt{\Delta t} \, g(T_n, y_{n-\tau}) \, \xi_{n+1} \\
\overset{\Delta t = 1}{\Leftrightarrow} \quad T_{n+1} &= f(T_n, y_{n-\tau}) + g(T_n, y_{n-\tau}) \, \xi_{n+1} \,. \label{eq:discrete_langevin}
\end{align}
Here, we set $\Delta t = 1$ and absorb the current state $T_n$ into the force term. Reminiscent of the continuous-time Langevin equation we refer to $f(T_n, y_{n-\tau})$ as drift and to $g(T_n, y_{n-\tau})$ as diffusion, both bearing the unit $[f]=[g]=[K]$. Here, $f(T_n, y_{n-\tau})$ and $g(T_n, y_{n-\tau})$ are allowed to be nonlinear resulting in a nonlinear restoring force and multiplicative noise, $\xi_n$ denotes Gaussian white noise with $\langle \xi_n \rangle = 0$ and $\langle \xi_n \xi_{n'} \rangle = \delta_{nn'}$. We assume $g(T,y)\geq 0 \; \forall T,y$. %Because of its resemblance to the continuous-time Langevin equation, we refer to Eq.~\eqref{eq:discrete_langevin} as the discrete-time Langevin equation. 
Equation~\ref{eq:discrete_langevin} defines a Markov chain with a continuous state-space.

Merging the terms of the continuous-time Langevin equation as above, and making ansatzes $\Phi(T_n, y_{n-\tau}; \{\lambda\})$ for the drift term and $\Theta(T_n, y_{n-\tau}; \{\theta\})$ for the squared diffusion term, we obtain the following likelihood function for the stochastic difference equation:
\begin{align}
\ln P(\{\lambda\}, \{\theta\}) = -\frac{1}{2}\sum_{i=1}^{N-1} \frac{(T_{i+1} - \Phi(T_i, y_{i-\tau}; \{\lambda\}))^2}{\Theta(T_i, y_{i-\tau}; \{\theta\})} - \sum_{i=1}^{N} \ln \sqrt{\Theta(T_i, y_{i-\tau}; \{\theta\})} \,,
\end{align}
in which $N$ is the length of the time series. We then obtain the optimal via the maximum likelihood, i.e. the minimization of the negative log-likelihood.
\begin{align}
    \{\hat{\lambda}\}, \{\hat{\theta}\} = \underset{\{\hat{\lambda}\}, \{\hat{\theta}\}}{\arg\min} \left[-\ln P(\{\lambda\}, \{\theta\})\right] \,.
\end{align}
This MLE is a modified least-squares fit in which the squares are weighted by the diffusion term $\Theta(T_n, y_{n-\tau}; \{\theta\})$.
Here, however, we continue by providing another, numerically cheaper approach.

Our proposed estimation procedure has two steps. In the first step, we estimate the deterministic force term $f$, followed by the estimation of the diffusion term $g$. We start by considering the force term. The conditional expectation for the next state $T_{n+1}$ solely depends on the current state $(T_n, y_{n-\tau})$ because of the Markovianity of the process:
\begin{align}
\langle T_{n+1} ~|~ T_n = T^*, y_{n-\tau} = y^* \rangle &= \langle f(T_n, y_{n-\tau}) + g(T_n, y_{n-\tau}) \, \xi_t ~|~ T_n = T^*, y_{n-\tau} = y^* \rangle \\ &= \langle f(x_t) ~|~ T_n = T^*, y_{n-\tau} = y^* \rangle + \langle g(T_n, y_{n-\tau}) \, \xi_t ~|~ T_n = T^*, , y_{n-\tau} = y^* \rangle \\ &= f(T^*, y^*) + \underbrace{\langle \xi_t \rangle}_{= 0} \; \langle g(T_n, y_{n-\tau}) ~|~ T_n = T^*, y_{n-\tau} = y^*  \rangle \\ &= f(T^*, y^*)
\end{align}
Here, we used the additivity of the conditional expectation, the independence of the noise and the diffusion term, and the vanishing mean of the noise. Hence, a single trajectory acts as a pseudo-ensemble: Averaging over many points of a time series in the vicinity %$\mathcal{U}(x^\prime)$
of a point $(T^*, y^*)$ we are left with the deterministic force.

For a given time series, we make an ansatz $\Phi(T,y; \{\lambda\})$ for the drift $f(T,y)$. The functional form of $\Phi$ requires an educated guess upon inspection of the averaged drift terms in the $(T_{n+1}; T_n, y_{n-\tau})$ space. %Demanding stability of the process requires $f(T,y)$ to not cross the identity, $|f(T,y)| < |T| \; \forall \; (T,y)$ \cite{Tong1993}.
We then find the optimal parameters $\{\hat{\lambda}\}$ by a least-squares fit of the averaged force terms in bins $\{B\}$, i.e.
\begin{align}
\{\hat{\lambda}\} = \underset{\{\lambda\}}{\mathrm{arg}\min} \sum_{j=1}^{|\{B\}|} \left(\left\langle T_{n+1}|(T_n, y_{n-\tau}) \in B_j\right\rangle - \Phi(T_j, y_j; \{\lambda\})\right)^2
\end{align}
in which $(T_j, y_j)$ is the center point of bin $B_j$. For a drift function $\Phi\left(T, y; \{\hat{\lambda}\}\right)$ which resembles $f(T,y)$, the residuals $R_n = T_{n+1} - \Phi\left(T_n, y_{n-\tau}; \{\hat{\lambda}\}\right)$ amount to the noise term in the stochastic difference equation (cf. Eq.~\eqref{eq:discrete_langevin}):
\begin{align}
R_n = g(T_n, y_{n-\tau}) \, \xi_{n+1} \,.
\end{align}

Assuming the optimal $\Phi$ matches the drift term $f$, we continue by determining the noise term $g$. Considering the conditional expectation of the squared residuals $R^2_n$, we obtain
\begin{align*}
\langle {R_n}^2 | T_n = T^*, y_{n-\tau} = y^* \rangle &= \langle g(T_n, y_{n-\tau})^2 \,\xi_n^2 | T_n = T^*, y_{n-\tau} = y^* \rangle \\ &= \langle \xi_n^2 \rangle \langle g(T_n, y_{n-\tau})^2 | T_n = T^*, y_{n-\tau} = y^* \rangle  \\ &= \underbrace{ \sigma^2}_{= 1} \, \langle g(T_n, y_{n-\tau})^2 | T_n = T^*, y_{n-\tau} = y^* \rangle \\
&= g(T_n, y_{n-\tau})^2 \,.
\end{align*}
Here, we used the independence of the noise $\xi_{n+1}$ of $g(T_n, y_{n-\tau})$ and $T_n$. Once again, by averaging over many data points of the time series in the vicinity of a sampling point $(T^*, y^*)$, the variance of the residuals is given by the squared diffusion term. Hence, in order to obtain a parametrized diffusion estimate, we make an ansatz $\Theta(T, y; \{\theta\})$ for the bin averages of the squared residuals. For this, we make an educated guess for its functional form based on the inspection of the bin averages in the $(R_n^2, T_n, y_{n-\tau})$ space.

Performing a least-squares fit yields the optimal parameters for approximating $g^2$:
\begin{align}
\hat{\theta} = \underset{\{\theta\}}{\mathrm{arg}\min} \sum_{j=1}^{|\{B\}|} \left(\left\langle R_n^2\left(\hat{\lambda}\right) | (T_n, y_{n-\tau}) \in B_j \right\rangle - \Theta\left(T_j, y_j; \{\theta\}\right)\right)^2 \,. %= \underset{\{\theta\}}{\mathrm{arg}\min} \, S_2(\theta) \,.
\end{align}
Once again, $(T_j, y_j)$ denotes the center point of bin $B_j$.
Although the binning procedure is formally required to obtain the drift and diffusion estimates, we apply the least-squares estimation without prior binning and obtain similar results for $\Phi$ and $\Theta$.

To ensure stability in cases of space-dependent diffusion, i.e. multiplicative noise, we impose bounds on the diffusion term such that the diffusion saturates at the boundaries of the observed data:
\begin{align*}
\tilde{\Theta}(T, y; \hat{\theta}) &= \Theta(\tilde{T}, \tilde{y}; \hat{\theta}) \\
\tilde{T} &= \begin{cases} T_{\mathrm{max}} \quad \mathrm{if} \quad T > T_{\mathrm{max}} \,, \\
T_{\mathrm{min}} \quad \mathrm{if} \quad T < T_{\mathrm{min}} \,,\\
T \quad \mathrm{else} \,. \end{cases} \\
\tilde{y} &= \begin{cases} y_{\mathrm{max}} \quad \mathrm{if} \quad y > y_{\mathrm{max}} \,, \\
y_{\mathrm{min}} \quad \mathrm{if} \quad y < y_{\mathrm{min}} \,, \\
y \quad \mathrm{else} \,. \end{cases}
\end{align*}
Then, the binding deterministic force term guarantees stability.

Choosing appropriate functions $\Phi$ and $\Theta$ is crucial for obtaining a suitable model. Therefore, we advise testing various functions and base the selection both on goodness of fit as well as comparisons of model data and original data.

%	\newpage
\bibliography{winter_extreme_temp}